\documentclass[twocolumn,showpacs,aps,pra,amsfonts,amsmath,amssymb,floatfix]
{revtex4-1}
\usepackage{subfigure}
\usepackage[]{graphicx, amsfonts, amsmath, amssymb, amstext, latexsym, float}
\usepackage{comment}
\usepackage{fullpage}

\begin{document}
\title{Capacity of optical communication in loss and noise with general quantum Gaussian receivers}

\date{\today}
\author{Masahiro Takeoka$^1$ and Saikat Guha$^2$ 
\\ $^1$ {\it National Institute of Information and Communications Technology, 
Koganei, Tokyo 184-8795, Japan}
\\ $^2$ {\it Quantum Information Processing Group, 
Raytheon BBN Technologies, Cambridge, MA 02138, USA}
}

\begin{abstract}
Laser-light (coherent-state) modulation is sufficient to achieve the ultimate (Holevo) capacity of classical communication over a lossy and noisy optical channel, but requires a receiver that jointly detects long modulated codewords with highly nonlinear quantum operations, which are near-impossible to realize using current technology. We analyze the capacity of the lossy-noisy optical channel when the transmitter uses coherent state modulation but the receiver is restricted to a general quantum-limited {\em Gaussian} receiver, i.e., one that may involve arbitrary combinations of Gaussian operations (passive linear optics: beamsplitters and phase-shifters, second order nonlinear optics (or active linear optics ): squeezers, along with homodyne or heterodyne detection measurements) and any amount of classical feedforward within the receiver. Under these assumptions, we show that the Gaussian receiver that attains the maximum mutual information is either homodyne detection, heterodyne detection, or time sharing between the two, depending upon the received power level. In other words, our result shows that to exceed the theoretical limit of conventional coherent optical communications, one has to incorporate non-Gaussian, i.e., third or higher-order nonlinear operations in the receiver. 
Finally we compare our Gaussian receiver limit with experimentally feasible non-Gaussian receivers and show that in the regime of low received photon flux, it is possible to overcome the Gaussian receiver limit by relatively simple non-Gaussian receivers based on photon counting. 

\end{abstract}



\maketitle

\section{Introduction}
\label{intro}

The bosonic channel plays a crucial role in classical and quantum information theory applied to optical communications. Classical capacity of the bosonic channel is particularly important since it determines the ultimate performance limit of conventional optical communication technology. Derivation of this ultimate limit, the Holevo capacity, is a nontrivial theoretical problem on which much effort has been spent~\cite{Caves1994,Yuen1992,Giovannetti2004,Giovannetti2004PRA,Llo10,Guh11,Wil12,Wil12a,Giovannetti2004PRA,Koenig2013,Giovannetti2013,Giovannetti_MOE2013}. If the channel is lossless and noiseless, capacity can be attained by modulating photon number states 
and an ideal photon number counting receiver~\cite{Yuen1992,Caves1994}. For a pure-loss channel, the capacity was found in~\cite{Giovannetti2004}: 
\begin{equation}
\label{eq:lossy_bosonic_capacity}
C_{\rm Loss} = g( \eta \bar{N} ),
\end{equation}
where $\eta$ is the channel's power transmissivity, $\bar{N}$ is the average input photon number per mode, and $g(x)=(x+1)\log(x+1) - x \log x$ 
(throughout the paper, we choose $\log \equiv \log_2$). 
It was also shown~\cite{Giovannetti2004} that capacity can be attained by a coherent-state modulation, but the optimal receiver must use joint measurements over many channel uses, which are very hard to realize physically~\cite{Llo10, Guh11, Wil12, Wil12a}. For lossy bosonic channel with additive thermal noise from the environment, it was conjectured~\cite{Giovannetti2004PRA} that the capacity is given by 
\begin{equation}
\label{eq:lossy_noisy_bosonic_capacity}
C_{\rm Thermal} = g( \eta \bar{N} + (1-\eta)N_{th} ) - g( (1-\eta)N_{th} ),
\end{equation}
where $N_{th}$ is the average number of noise photons per mode. The above conjecture relied on an unproven minimum output entropy conjecture, which was recently proven~\cite{Giovannetti_MOE2013} to be true, hence confirming that~\eqref{eq:lossy_noisy_bosonic_capacity} indeed is the capacity of the lossy-noisy bosonic channel, and that it can be achieved using a coherent-state modulation.

One of the practically important things to note is that the capacities $C_{\rm Loss}$, and $C_{\rm Thermal}$ can be achieved by encoding information in coherent states with a Gaussian distribution, i.e., one does not need to use nonclassical states such as entangled or squeezed states. Unlike the simple transmitter, the receiver must use a decoding strategy that is highly non-classical. In the proof of achievability in the original HSW theorem~\cite{Holevo1998,Hausladen1996,Schumacher1997}, they employed the {\em square-root measurement} (SRM) over an infinite sequence of signal states 
(codeword). This is in general a {\em collective} measurement which may include entangling operations between modulation symbols across channel uses and thus is a highly nonclassical operation. Since there is a clear gap between the Holevo capacity $C_{\rm Loss}$ and the Shannon capacities of conventional optical receivers~\cite{Giovannetti2004,Wil12}, i.e., homodyne, heterodyne or direct-detection receivers, a natural question arises: how to design receiver measurements that can achieve a reliable communication rate beyond the Shannon limits of conventional receivers using a receiver that is more practical than the SRM, the mathematical specification of which unfortunately sheds no light on a structured receiver design? This question has been explored theoretically~\cite{Sasaki1997,Sasaki1998,Buck2000,Guh11} and experimentally \cite{Fujiwara2003,Takeoka2004}, but a truly-realizable receiver specification that outperforms conventional optical receivers still eludes us. 

In this paper, we restrict ourselves to receiver measurements consisting of a limited class of quantum operations, called Gaussian operations, and allow any amount of classical feedback or feedforward (FF) within the receiver. This class of operations can be constructed by combining passive linear optics (a network of beamsplitters and phase-shifters), second order nonlinear optical processes (squeezers), along with homodyne and heterodyne detection measurements, all of which are now routinely implemented~\cite{Braunstein2005RMP,Weedbrook2012}. On the other hand, it has also been found that some of the important quantum protocols cannot be performed with Gaussian operations and classical processing alone. Examples of those are universal quantum computing~\cite{Bartlett2002}, entanglement distillation of Gaussian states~\cite{Eisert2002,Fiurasek2002,Giedke2002}, optimal cloning of coherent states~\cite{Cerf2005}, optimal discrimination of coherent states~\cite{Takeoka2008,Tsujino2011,Wittmann2010-1,Wittmann2010-2}, and Gaussian quantum error correction~\cite{Niset2009}. Here we ask the question whether the classical information transmission 
capacity for bosonic channels can benefit from a receiver that is restricted to Gaussian quantum operations and FF. The main result of our paper is an addition to the long list of the above ``no-go theorems'', i.e., we show that Gaussian operations and classical FF processing cannot 
exceed the Shannon limit of homodyne and heterodyne detection for the lossy-noisy bosonic channel 
with coherent state inputs. Our result has a practical importance: to exceed the theoretical limit of the conventional coherent optical communication system via a collective quantum decoder, one has to incorporate non-Gaussian operations in the receiver. As a byproduct of the analysis, we propose a hybrid receiver that time-shares between homodyne and heterodyne, which we show to slightly 
surpass the envelope of the homodyne and heterodyne capacities in a certain regime of average signal photon number per mode. 

Finally we compare our Gaussian receiver limit with 
experimentally feasible non-Gaussian receivers based on photon counting 
detectors. We show that for extremely low signal power, it is possible to 
overcome the Gaussian receiver limit via relatively simple 
non-Gaussian receivers.

\begin{figure}
\begin{center}
\includegraphics[width=0.6\linewidth]{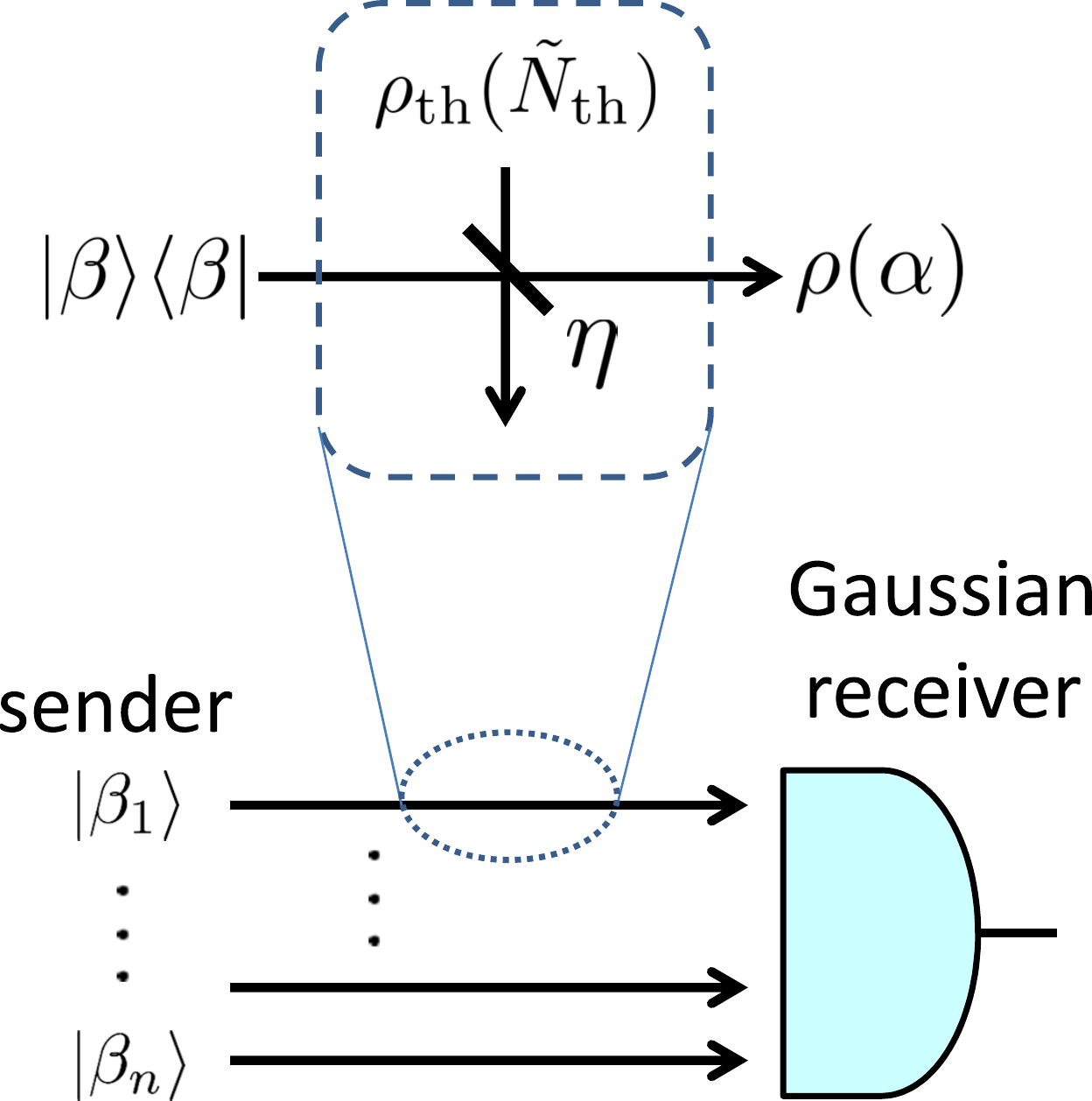}   %
\caption{\label{fig:lossy_noisy_bosonic_channel}
(Color online) Lossy and noisy bosonic channel model and $n$ use of 
the channels. 
}
\end{center}
\end{figure}

\section{Channel model}
\label{sec:channel_model}

The sender encodes messages in coherent states and sends them via multiple uses of a lossy-noisy bosonic channel $\mathcal{N}_{\rm B}$, each use of which can be modeled by a beam splitter with transmissivity $\eta \in (0, 1]$ with a zero-mean thermal state injected into the other input port (see Fig.~\ref{fig:lossy_noisy_bosonic_channel}). Let $\tilde{N}_{\rm th}$ be the mean photon number of the thermal state. For a coherent state input $|\beta\rangle$, $\beta \in {\mathbb C}$, the output is a displaced thermal state:
\begin{equation}
\label{eq:channel_output}
\mathcal{N}_{\rm B} (|\beta\rangle\langle\beta|) = 
D(\alpha) \rho_{\rm th}(N_{\rm th}) D^\dagger(\alpha) \equiv \rho(\alpha) 
\end{equation}
where $D(\alpha)=\exp[\alpha \hat{a}^\dagger - \alpha^* \hat{a}]$ is a displacement operation, $\hat{a}$ and $\hat{a}^\dagger$ are the annihilation and creation operators, respectively, $\rho_{\rm th}(N_{\rm th})$ is a thermal state with mean photon number $N_{\rm th}$, and
\begin{equation}
\label{eq:alpha_N_th}
\alpha = \sqrt{\eta}\beta , \quad 
N_{\rm th} = (1-\eta) \tilde{N}_{\rm th} .
\end{equation}

The sender prepares an $n$-mode coherent state,
$
\rho_S = \int d^{2n} \underline{\beta} \, P(\underline{\beta}) 
|\underline{\beta}\rangle\langle\underline{\beta}|,
$
to encode a message, where $|\underline{\beta}\rangle = |\beta_1\rangle \otimes \cdots \otimes |\beta_n\rangle$, with $\underline{\beta}=[\beta_1 , \cdots, \beta_n]^T \in {\mathbb C}^n$and $P(\underline{\beta})$ a probability distribution function. The modulation power constraint translates to: 
\begin{equation}
\label{eq:input_power_constraint}
\int d^{2n} \underline{\beta} \, P(\underline{\beta}) |\underline{\beta}|^2
\le \tilde{N} .
\end{equation}
The $n$ channel use transmission transforms $\rho_S$ to:
\begin{equation}
\label{eq:input_signal}
\rho_R = 
\int d^{2n} \underline{\alpha} \, P(\underline{\alpha}) 
\rho_r (\underline{\alpha}) ,
\end{equation}
where $\underline{\alpha} = [\alpha_1, \cdots, ,\alpha_n]^T 
= \sqrt{\eta} \underline{\beta}$ and 
$\rho_r(\underline{\alpha}) = 
\underline{D}(\underline{\alpha}) 
\rho_{\rm th}^{\otimes n}(N_{\rm th})
\underline{D}^\dagger(\underline{\alpha}) ,
$
with $\underline{D}(\underline{\alpha})=D(\alpha_1)\otimes
\cdots \otimes D(\alpha_n)$, 
\begin{equation}
\label{eq:output_power_constraint}
\int d^{2n} \underline{\alpha} \, P(\underline{\alpha}) |\underline{\alpha}|^2
\le \bar{N} ,
\end{equation}
and $\bar{N} = \eta \tilde{N}$. 
A {\em quantum Gaussian receiver} decodes the message from $\rho_R$, which in general could make a Gaussian collective measurement involving any Gaussian operation, classical FF and post processing. Let $\{ \Pi(\underline{\alpha}_M) \}$ be a positive operator-valued measure (POVM) for an $n$-mode quantum Gaussian receiver with measurement outcome $\underline{\alpha}_M$. The maximum reliable information rate per channel use is given by 
\begin{equation}
\label{eq:capacity_regularized}
C(\mathcal{N}) = \lim_{n \to \infty} 
\frac{1}{n} \max_{P,\,\Pi} I(X^n;Y^n) , 
\end{equation}
where $I(X^n;Y^n)$ is the mutual information calculated for priors $P_{X^n}(x^n) = P(\underline{\alpha})$ and 
transition probabilities, $
P_{Y^n|X^n}(y^n|x^n) = P(\underline{\alpha}_M|\underline{\alpha}) 
=
{\rm Tr}[ \rho_r(\underline{\alpha}) \Pi(\underline{\alpha}_M)]
$.

\section{Capacity of a lossy-noisy bosonic channel with coherent states and Gaussian receiver}
\label{sec:main}

\begin{figure}
\begin{center}
\includegraphics[width=1\linewidth]{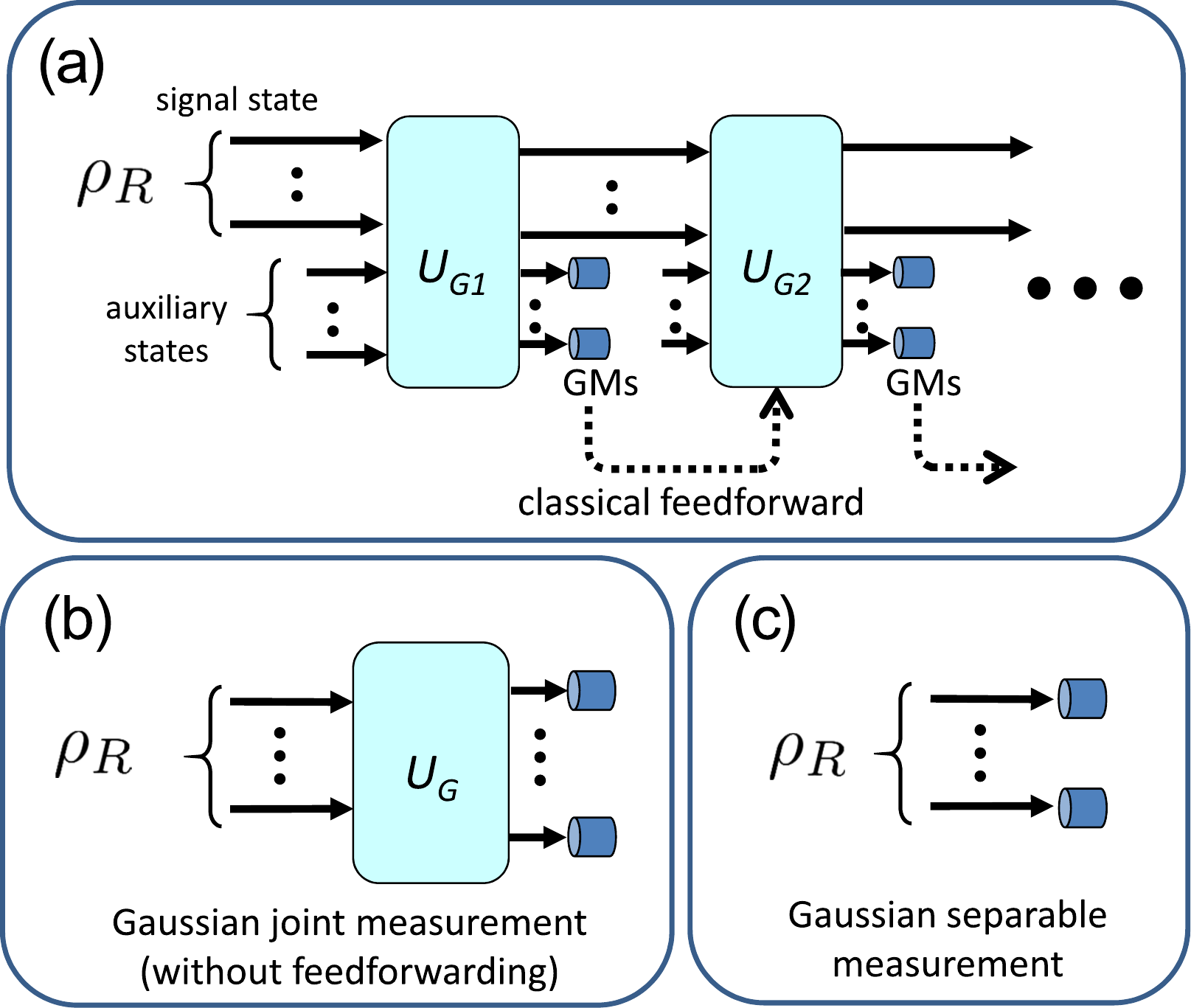}   %
\caption{\label{fig:general_gaussian}
(Color online) 
(a) General model of Gaussian receiver, 
(b) Gaussian joint measurement, and 
(c) Gaussian separable measurement. 
$U_G$: Gaussian unitary operation, GM: Gaussian measurement, 
$\rho_R$: signal state after transmitting 
the lossy and noisy bosonic channels. 
}
\end{center}
\end{figure}

In this section, we derive an explicit expression for Eq.~(\ref{eq:capacity_regularized}) and show that the optimal quantum Gaussian receiver is simply given by a homodyne or heterodyne receiver, or an appropriate time sharing between them. Let us consider the general structure of an $n$-mode Gaussian receiver. As illustrated in Fig.~\ref{fig:general_gaussian}(a), it can be decomposed into ($n+m$)-mode-input Gaussian unitary operations $U_{Gi}$, $i = 1, 2, \ldots$ with $m$-mode ancillary Gaussian input, single mode Gaussian measurements (without classical FF) and classical FF into subsequent unitaries. Without loss of generality, we consider only `noise-free' 
operations that map pure states into pure states (we can simulate any noisy process by simply discarding 
a part of the system in a noise-free operation \cite{Giedke2002}).

We prove the statement of optimality stated above, by showing:
(1) the classical FF is not necessary to maximize $I(X^n;Y^n)$, 
(2) the optimal Gaussian measurement is a separable one, and
(3) the optimal separable measurement is given by homodyne, heterodyne 
or time sharing between them.

{\bf Step 1: Classical feedforward operations}.
First we show that the classical FF operations in Fig.~\ref{fig:general_gaussian}(a) are not necessary. One is able to show this by slightly extending the theorem on Gaussian operations shown in~\cite{Eisert2002,Fiurasek2002,Giedke2002} as a part of the no-go theorem on Gaussian entanglement distillation via Gaussian local operations. Precisely, in \cite{Eisert2002,Fiurasek2002,Giedke2002} it was shown 
that for a Gaussian state input, any trace decreasing Gaussian operation (such as a partial measurement) can always be transformed into a trace preserving (deterministic) operation by adding an appropriate conditional displacement operation. It is thus clear that if the input in Fig.~\ref{fig:general_gaussian}(a) is a Gaussian state, the conditional classical FF based on partial measurement outcomes are unnecessary. Note however that the received signal ensemble (Eq.~(\ref{eq:input_signal})) is a convex combination of displaced thermal states which in general could be a non-Gaussian state. Nevertheless, by slightly extending the above theorem, we can show that for any possible convex combination (i.e. for any probability distribution $P(\underline{\alpha})$), any trace decreasing Gaussian operation can be transformed into a trace preserving Gaussian operation. This allows us to conclude that the conditional operations with partial measurement and classical FF are not necessary in our receiver. 
Though the extension of \cite{Eisert2002,Fiurasek2002,Giedke2002} is rather straightforward, we describe it in Appendix \ref{appendixB} for completeness.

{\bf Step 2: Joint and separable measurement.}
Removing the partial measurements and classical FF, the receiver measurement is now as illustrated in Fig.~\ref{fig:general_gaussian}(b) where an $n$-mode Gaussian unitary operation $U_G$ is followed by $n$ heterodyne measurements. We now show that even a collective measurement is not necessary to obtain the maximum mutual information. Such a collective Gaussian detection is described 
by a set of operators 
$\{\Pi_G(\Gamma_{\mathcal{M}}, d_{\mathcal{M}})\}_{d_\mathcal{M}}$ 
where $\Gamma_{\mathcal{M}}$ and $d_{\mathcal{M}}$ are, respectively, 
the covariance matrix (CM) and displacement vector (DV) of the characteristic function: 
\begin{eqnarray}
\label{eq:GaussianMeas}
\chi_G (x) & = & {\rm Tr}\left[ 
\Pi_G(\Gamma_\mathcal{M}, d_\mathcal{M}) 
\mathcal{W}(x) \right]
\nonumber\\ & = & 
\exp\left[ -\frac{1}{4} x^T \Gamma_\mathcal{M} x 
+ i d_\mathcal{M}^T x \right] ,
\end{eqnarray}
where $\mathcal{W}(x)=\exp[-i x^T R]$ is the Weyl operator, $x\in\mathbb{R}^{2n}$, and $R=[ \hat{x}_1, \cdots , \hat{x}_n , \hat{p}_1, \cdots , \hat{p}_n]^T$ consists of quadrature operators satisfying commutation relation $[\hat{x}_k, \hat{p}_l]=i\delta_{kl}$. Note that the displacement vector $d_\mathcal{M}$ corresponds to a set of measurement outcomes. We summarize some basic properties of characteristic functions in the Appendix \ref{appendixA}.

\begin{figure}
\begin{center}
\includegraphics[width=0.5\linewidth]{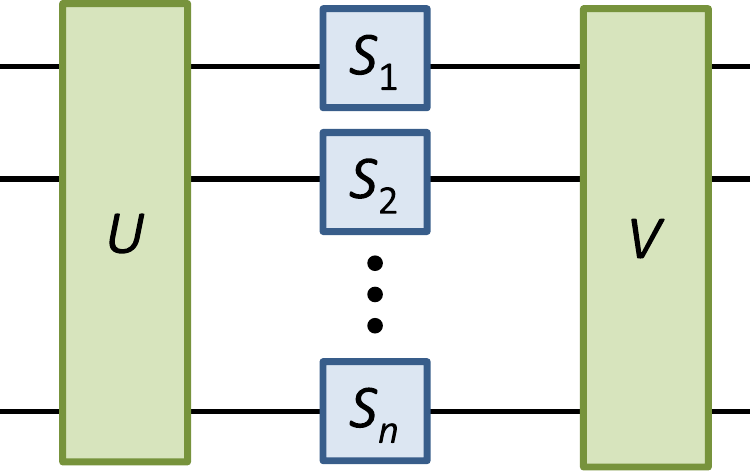}   %
\caption{\label{fig:gaussian_unitary}(Color online) 
Decomposition of Gaussian unitary operation. 
$U$, $V$: unitary operations with linear optics, 
$S_i$: single mode squeezers. 
}
\end{center}
\end{figure}

It is known that a Gaussian unitary operation can be decomposed into a passive linear optical unitary operation $U$ (implementable via a network of beamsplitters and phase-shifters), a set of single mode squeezers, and another passive linear optic unitary $V$ \cite{Braunstein2005} as illustrated in Fig.~\ref{fig:gaussian_unitary} (see also Eq.~(\ref{eq:S_decomposition}) in Appendix \ref{appendixA}). The covariance matrix and the displacement vector for the Gaussian measurement is now constructed as follows. A set of $n$ heterodyne measurements on $n$ modes is given by a correlation matrix $\gamma_{M_0} = I_{2n}$ and a $2n$ real vector $d_{M_0}$, where $I_{2n}$ is a $2n \times 2n$ identity matrix. Denoting the symplectic matrix for the linear operation $V$ as $S_V$, the CM and the DV of the measurement consisting of $V$ and 
the heterodyne receivers are $S_V^T \gamma_{M_0} S_V = I_{2n}$ and $S_V^T d_{M_0}$, respectively. 
Including the single-mode squeezers, the measurement is described by, $S_S^T S_V^T \gamma_{M_0} S_V S_S = \gamma_{\bar{M}} = {\rm diag}[e^{-2 r_1} \, \cdots \, e^{-2 r_n} \, e^{2 r_1} \, \cdots \, e^{2 r_n}]$ and $S_S^T S_V^T d_{M_0} = d_{\bar{M}}$, where $r_i$ are the squeezing parameters. Finally, adding the linear operation $U$ and defining $S_U^T d_{\bar{M}} \equiv d_{\tilde{M}}$, the characteristic function of the joint Gaussian measurement is given by: 
\begin{equation}
\label{eq:GaussianMeas}
\chi_G (x) = \exp\left[ -\frac{1}{4} x^T S_U^T \gamma_{\bar{M}} S_U x 
+ i d_{\tilde{M}}^T x \right] .
\end{equation}
Note that the entries of $d_{\tilde{M}}$ can be obtained by applying the linear transformation $V$, in software, on the measurement outcome $d_{M_0}$. Since this operation can be performed {\it after} the measurement we can remove $V$ from the Gaussian unitary operation without loss of generality.

The received ensemble is a set of displaced thermal states~(\ref{eq:input_signal}). In terms of the characteristic functions, denoting $\underline{\alpha}= [\alpha_1, \cdots, \alpha_n]^T$ 
by a 2$n$-length real vector $d_r = \sqrt{2}[-{\rm Im}\alpha_1, \cdots, -{\rm Im}\alpha_n, {\rm Re}\alpha_1, \cdots, {\rm Re}\alpha_n]^T$, the characteristic function of the displaced thermal state $\rho_r(\underline{\alpha})$ is given by 
\begin{equation}
\label{eq:ch_func_displ_thermal_state}
\chi_r (x) = \exp\left[ -\frac{1}{4} x^T \gamma_{\rm th} x + 
i d_r^T x \right] ,
\end{equation}
where,
\begin{equation}
\label{eq:displaced_thermal_cov_matrix}
\gamma_{\rm th} = \left[
\begin{array}{cc}
1+2N_{\rm th} & 0 \\
0 & 1+2N_{\rm th} 
\end{array}
\right]^{\oplus n} .
\end{equation}

The conditional probability of obtaining the outcome $d_{\tilde{M}}$ by detecting $\rho_r(\underline{\alpha})$ with the Gaussian measurement in Eq.~(\ref{eq:GaussianMeas})
is then given by 
\begin{eqnarray}
\label{eq:Joint_conditional_probability}
&& P(d_{\tilde{M}}|d_S) = \left( \frac{1}{2\pi} \right)^{2n} 
\int dx \, \chi_r(x) \chi_M(-x) 
\nonumber\\ && = \frac{1}{\sqrt{\det\left(\gamma_{\rm th} 
+ S_U^T \gamma_{\tilde{M}} S_U \right)}} 
\nonumber\\ && \times 
\exp\left[ -(d_S - d_{\tilde{M}})^T 
\frac{1}{\gamma_{\rm th} + S_U^T \gamma_{\tilde{M}} S_U } 
(d_S - d_{\tilde{M}}) \right] .
\nonumber\\
\end{eqnarray}
The expression in Eq.~(\ref{eq:Joint_conditional_probability}) is equivalent to that of a classical multi-dimensional Gaussian channel with a correlated noise CM, $\gamma_{\rm th} + S_U^T \gamma_{\bar{M}} S_U$~\cite{CoverThomas}. It is well known that the mutual information of a Gaussian channel is maximized by a Gaussian input distribution \cite{CoverThomas},
\begin{equation}
\label{eq:prior_probability}
P(d_r) = \frac{1}{2\pi^n \sqrt{\det P}} 
\exp\left[ - \frac{1}{2} d_r^T \frac{1}{P} d_r \right] ,
\end{equation}
where $P$ is a diagonal matrix \cite{comment_P} with the power constraint:
$
\frac{1}{2n}\sum_{i=1}^{2n} P_{ii} \le \bar{N}
$.
As a consequence the mutual information per channel use is given by 
$I(X;Y) = I(X^n;Y^n)/n$ where 
\begin{eqnarray}
\label{eq:Joint_Mutualinfo}
I(X^n;Y^n) & = & \frac{1}{2} \log \frac{ 
{\rm det}(2P + \gamma_{\rm th} + S_U^T \gamma_{\bar{M}} S_U) }{
{\rm det}(\gamma_{\rm th} + S_U^T \gamma_{\bar{M}} S_U) }
\nonumber\\ & = & 
\frac{1}{2} \log \frac{ 
{\rm det}(2P + \gamma_{\rm th} + S_U^T \gamma_{\bar{M}} S_U) }{
{\rm det}(\gamma_{\rm th} + \gamma_{\bar{M}}) } .
\nonumber\\
\end{eqnarray}
The second equality follows from: $\gamma_{\rm th} = (1+2N_{\rm th})I_{2n}$ and unitarity of $S_U$. To maximize $I(X^n;Y^n)$, we need to optimize $S_U$ such that 
${\rm det}(2P + \gamma_{\rm th} + S_U^T \gamma_{\bar{M}} S_U)$ 
is maximized.
According to the Hadamard inequality: 
\begin{equation}
\label{eq:Hadamard_inequality}
{\rm det} \left( X \right) \le \prod_i X_{ii} ,
\end{equation}
where $X$ is a positive matrix and equality holds iff 
$X$ is diagonal, the maximum is obtained when 
$2P + \gamma_{\rm th} + S_U^T \gamma_{\bar{M}} S_U$ is diagonal.
Since $P$ and $\gamma_{\rm th}$ are diagonal 
and the trace of $S_U^T \gamma_{\bar{M}} S_U$ 
is invariant under any $S_U$, we conclude that 
$S_U=I_{2n}$ is optimal. 
As a consequence, the passive-linear-optic unitary $U$ is unnecessary. 
Since we have already concluded that the other unitary $V$ is also unnecessary, we conclude that it is sufficient for the optimal quantum Gaussian receiver to make a set of separable measurements (Fig.~\ref{fig:general_gaussian}(c)).

\begin{figure}[htbp]
\centering
\subfigure
{
\includegraphics[width=1\linewidth]
{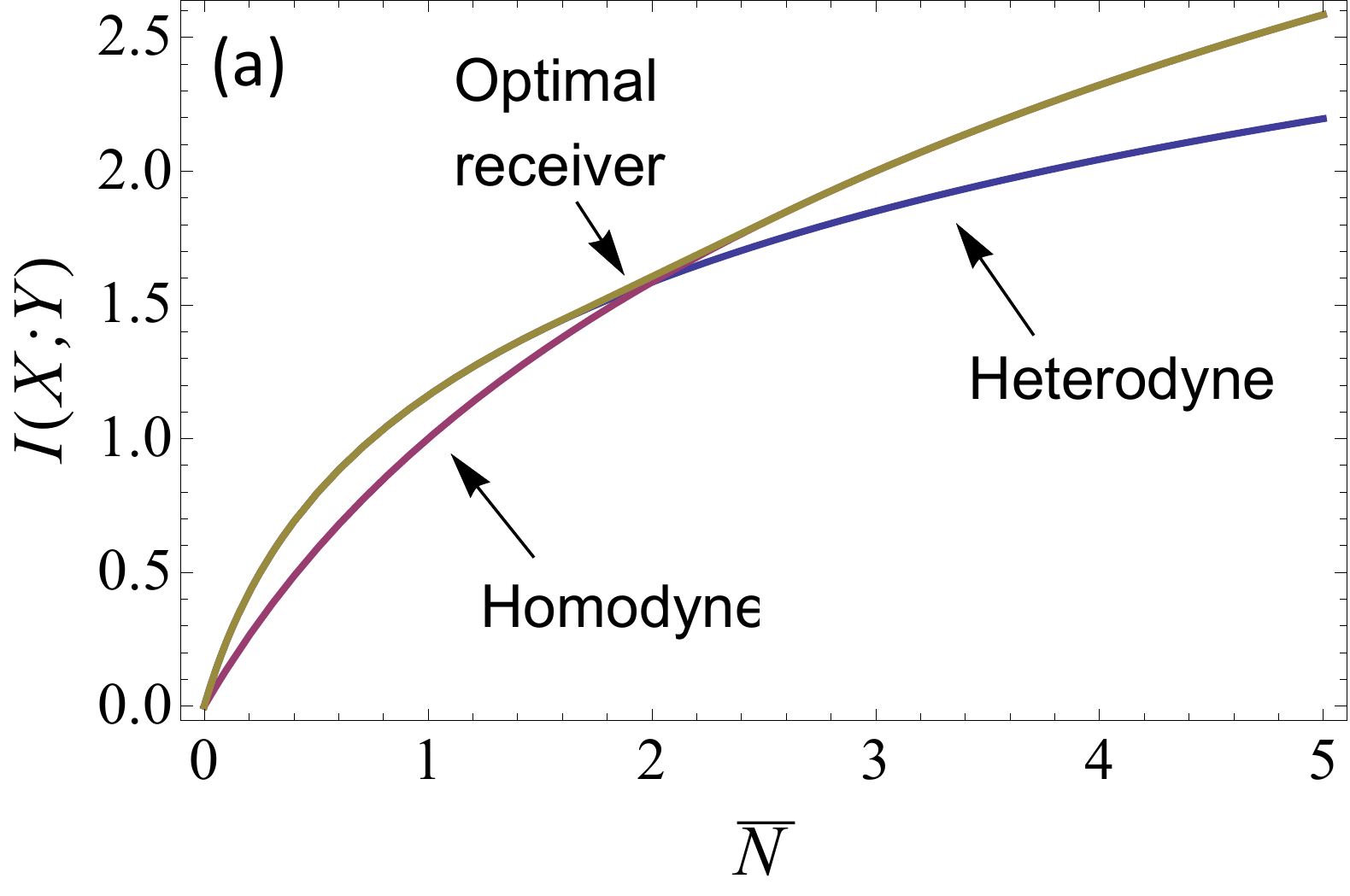}
\label{Fig4a}
}
\subfigure
{
\includegraphics[width=1\linewidth]
{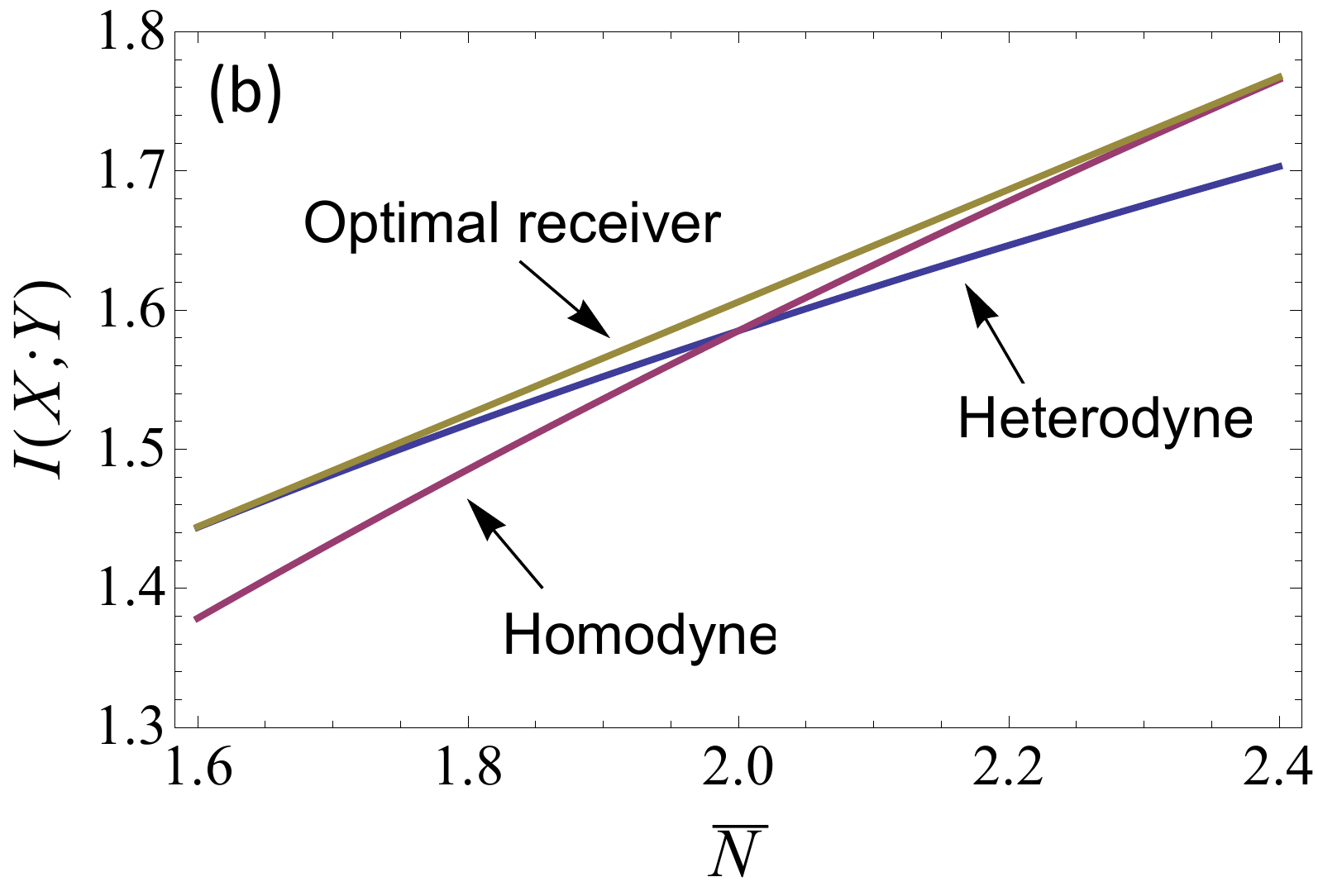}
\label{Fig4b}
}
\caption{
\label{fig:mutual_information}(Color online) 
Capacity of homodyne (violet), heterodyne (blue), 
and optimal (yellow) receivers for a pure-loss optical channel 
as a function of the average photon number at the receiver 
$\bar{N} = \eta \tilde{N}$. (a) and (b) show the same plots 
with different $x$-$y$ ranges. 
}
\end{figure}

\begin{figure}
\begin{center}
\includegraphics[width=1\linewidth]{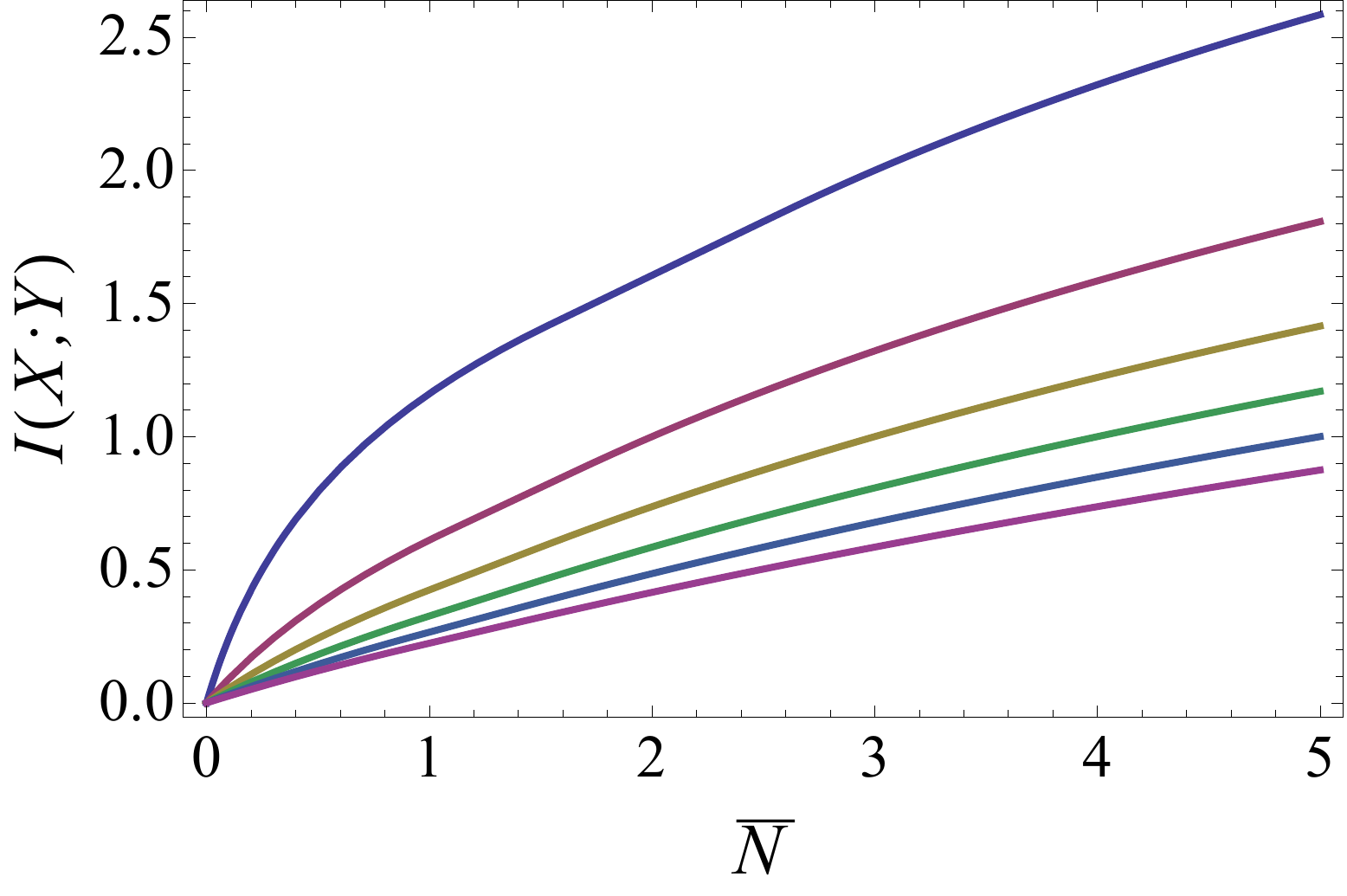}   %
\caption{\label{fig:mutual_info_thermal}(Color online) 
Cpacity of optimal Gaussian receiver for optical channels 
with loss and thermal noise. The solid lines from top to bottom 
correspond to $N_{\rm th}=0,\,1,\,2,\,3,\,4,\,5$. 
}
\end{center}
\end{figure}

{\bf Step 3: Optimization of the separable receiver.} We split this part into the following two steps. We first consider a fixed receiver for all $n$ channel uses and show that the optimal measurement is given either by a homodyne or a heterodyne measurement depending on the value of $\bar{N}$. Next we show that in a given range of $\bar{N}$ values, one can further optimize the mutual information by sharing the channel uses between homodyne and heterodyne measurements with an optimal power allocation across the channel uses.

As a first step, let us consider the maximization of the single mode mutual information
\begin{equation}
\label{eq:single-mode_mutual_info}
I(X;Y) = \frac{1}{2} \log \frac{{\rm det}(2P^{(1)} + \gamma_{\rm th}^{(1)} 
+ \gamma_M^{(1)}) }{
{\rm det}( \gamma_{\rm th}^{(1)} + \gamma_M^{(1)} ) },
\end{equation}
by optimizing a single-mode measurement $\gamma_M^{(1)}$ and power distribution $P^{(1)}$ (the superscripts denote $n=1$). As mentioned above, $I(X;Y)$ is maximized with diagonal $P^{(1)}$ and $\gamma_M^{(1)}$. General expressions of diagonal $P^{(1)}$ and $\gamma_M^{(1)}$ are given by 
\begin{equation}
\label{eq:single-mode_correlation_matrices}
P^{(1)} = \left[
\begin{array}{cc}
N_1 & 0 \\
0 & N_2 
\end{array}
\right] , \qquad 
\gamma_M^{(1)} = \left[
\begin{array}{cc}
e^{-2r} & 0 \\
0 & e^{2r} 
\end{array}
\right] , 
\end{equation}
where the power constraint is 
$(N_1 + N_2)/2 = \bar{N}$. 
Note that $r=\pm\infty$ and $r=0$ correspond to homodyne and heterodyne 
measurement, respectively. 
Substituting Eq.~(\ref{eq:single-mode_correlation_matrices}) 
and $N_2 = 2\bar{N} - N_1$ into Eq.~(\ref{eq:single-mode_mutual_info}), 
we have 
\begin{eqnarray}
\label{eq:single-mode_mutual_info2}
I(X;Y) & = & 
\frac{1}{2} \log \left[\frac{ 
(2N_1 + 2N_{\rm th} + e^{-2r}) }{ 
( 1 + 2N_{\rm th} + e^{-2r} ) } \right. 
\nonumber\\ && \times \left. 
\frac{(4\bar{N} - 2N_1 + 2N_{\rm th} + e^{2r})}{
( 1 + 2N_{\rm th} + e^{2r} )} \right] 
 \equiv f (N_1, r), \nonumber
\end{eqnarray}
which we want to maximize over $0 \le N_1 \le 2\bar{N}$ and 
$r \in (-\infty, \infty)$. By evaluating $\partial f(N_1, r)/\partial N_1$ and 
$\partial f(N_1, r)/\partial r$, we find that the extremum could exist 
only at $(N_1, r) = (\bar{N}, 0)$ with 
$
f(\bar{N}, 0) = \log \left(({1+N_{\rm th}+\bar{N}})/({1+N_{\rm th}})\right).
$
On the other hand, for $r \to \pm \infty$ or 
$N_1 = 0, 2\bar{N}$, the maximum $f$ is obtained at 
$(N_1, r) = (2\bar{N}, \infty)$ and $(0, -\infty)$ with, 
\begin{equation}
f(2\bar{N}, \infty) = f(0, -\infty) = \frac{1}{2} 
\log \frac{1+2N_{\rm th}+4\bar{N}}{1+2N_{\rm th}}.
\end{equation}
Therefore, the maximum mutual information is given by 
\begin{eqnarray}
\label{eq:single-mode_mutual_info3}
I(X;Y) & = & \max \left[ 
\frac{1}{2} \log \frac{1+2N_{\rm th}+4\bar{N}}{1+2N_{\rm th}},
\right. 
\nonumber\\ && \left. 
\log \frac{1+N_{\rm th}+\bar{N}}{1+N_{\rm th}} \right], 
\nonumber\\ & = & \left\{
\begin{array}{ll}
\frac{1}{2} \log \frac{1+2N_{\rm th}+4\bar{N}}{1+2N_{\rm th}} & 
0\le\bar{N}\le\frac{2(1+N_{\rm th})}{1+2N_{\rm th}} \\
\log \frac{1+N_{\rm th}+\bar{N}}{1+N_{\rm th}} & 
\bar{N}\ge\frac{2(1+N_{\rm th})}{1+2N_{\rm th}}
\end{array} \right. ,
\nonumber\\
\end{eqnarray}
which implies that if we fix the measurement on each mode, homodyne or heterodyne measurement is optimal. In other words, squeezing cannot increase the mutual information. 

The above result is slightly improved around $\bar{N}=2(1+N_{\rm th})/(1+2N_{\rm th})$ by optimizing the power allocation between channel uses. 
In the following, to simplify the expressions, we set $N_{\rm th}=0$ 
although the same approach works for finite $N_{\rm th}$.

Consider $n$ channel uses (modes) and suppose homodyne detection is used on the first $t$ uses and heterodyne detection on the rest. We can optimize the power allocation for these $n$ mode under the condition $\sum_i N_i \le n \bar{N}$ where $N_i$ is the average photon number of the $i$th mode. This optimization can be carried out by the Lagrange multiplier method. Defining
\begin{eqnarray}
\label{eq:Lagrange_multipliers}
&& F(N_1, \cdots, N_n) = \sum_{i=1}^t \frac{1}{2} \log \left( 1+4N_i \right)
\nonumber\\ &&
+ \sum_{i=t+1}^n \log \left( 1+N_i \right) 
+ \lambda \left( \sum_{i=1}^n N_i - \bar{N} \right) ,
\end{eqnarray}
where $\lambda$ is a Lagrange multiplier. Solving $(d F)/(d N_i) = 0$ for $i=1, \cdots, n$, we find the optimal photon numbers, $N_i$ as
\begin{eqnarray}
\label{eq:N_i_optimal}
N_i & = & \frac{1}{4}( \nu -1 )  \quad (1 \le i \le t) , 
\\
N_i & = & \frac{\nu}{2} - 1 \quad (t < i \le n) ,
\end{eqnarray}
where $\nu = (4\bar{N}+1+3x)/(1+x)$ and 
$x=(n-t)/n$.
The mutual information is then given by 
\begin{equation}
\label{eq:N_i_optimized_mutual_info}
\frac{1}{n} I(X^n; Y^n) = \frac{1}{2} (1+x) \log \frac{4\bar{N} + 1 + 3x}{1+x} 
-x.
\end{equation}
We can futher optimize $x$ (or equivalently $\nu$) 
and obtain 
\begin{equation}
\label{eq:x_optimized_mutual_info}
\frac{1}{n} I(X^n; Y^n) = 
\frac{1}{\nu^* - 3}\left( \log \nu^* -2 \right) \left( 2 \bar{N} - 1 \right) 
+1 ,
\end{equation}
where $\nu^*$ is one of the solutions of 
$
\nu (1+ 2 \ln 2 -\ln \nu) = 3
$,
which satisfies $N_i >0$ for all $i$.  
Numerically, this is $\nu^* = 7.145\cdots$. 
The optimal $x$ is then given in terms of $\nu^*$ as 
$x = ({4\bar{N} +1 - \nu^*})/({\nu^*-3})$,
which yields $0 < x \le 1$ when 
$(\nu^* -1)/4 < \bar{N} \le (\nu^*-2)/2$. 
Therefore, in summary, the capacity of the pure-loss optical channel 
with laser-light modulation and a general quantum Gaussian receiver is given by 
\begin{widetext}
\begin{equation}
\label{eq:gaussian_capacity}
C = \left\{ 
\begin{array}{lll}
\frac{1}{2} \log (1+4\bar{N}) & \quad 0 \le \bar{N} \le \frac{\nu^*-1}{4} 
& {\rm (homodyne)} \\
\frac{\log \nu^* -2}{\nu^*-3} \left( 2\bar{N} - 1 \right) + 1
& \quad \frac{\nu^*-1}{4} < \bar{N} \le \frac{\nu^*-2}{2} 
& {\rm (homodyne+heterodyne)} \\
\log ( 1 + \bar{N} ) & \quad  \frac{\nu^*-2}{2} < \bar{N} 
& {\rm (heterodyne)} 
\end{array}
\right. ,
\end{equation}
\end{widetext}
where $(\nu^*-1)/4 = 1.536\cdots$ and $(\nu^*-2)/2 = 2.572\cdots$. 
This is illustrated in Fig.~\ref{fig:mutual_information}. 

The capacities for the channel with finite $N_{\rm th}$ 
can be derived from Eq.~(\ref{eq:single-mode_mutual_info3})
by the same optimization procedure. 
We plot the numerical results in Fig.~\ref{fig:mutual_info_thermal}.

\section{Non-Gaussian receivers based on photon counting}
\label{sec:non-gaussian}

In the previous sections, we showed that one has to incorporate non-Gaussian operations in the receiver in order to exceed the theoretical limit of conventional coherent optical communication. In this section, we compare that limit with the performance theoretically achievable using currently known structured non-Gaussian optical receivers, and also with the ultimate (Holevo) capacity limit. We only consider separable non-Gaussian receivers---i.e., receivers that detect each modulation symbol one at a time. One of the practical, but highly non-Gaussian, operations for an optical communication receiver is photon counting. This falls under the category of {\em direct detection} receivers (unlike homodyne and heterodyne detection, which are collectively fall under {\em coherent detection} receivers). Direct detection receivers include the photon number resolving detector (PNRD) and the on-off single-photon detection (SPD) receiver, where the latter can discriminate only between zero and non-zero photons. 

It has been shown theoretically that combining an ideal single-photon detector with the phase-space displacement operation (implementable using a highly-transmissive beamsplitter and a strong coherent-state local oscillator), and potentially also classical feedback (e.g., subsequent single-photon detection events triggering real-time updates to the amplitude and phase of a local oscillator that in turn is mixed with, to coherently {\em null}, the received pulse before it is incident on the detector's active surface) can beat the coherent-detection (homodyne and heterodyne) limit of discriminating two or more coherent states 
\cite{Takeoka2008,Kennedy1973,Dolinar1973,Bondurant1993,Silva2013}. This 
was experimentally verified, without correcting for imperfections, for the binary phase-shift keyed (BPSK) signal ($\left\{|\alpha\rangle, |-\alpha\rangle\right\}$)~\cite{Tsujino2011} and for the quadrature phase-shift keyed (QPSK) signal ($\left\{|\alpha\rangle, |i\alpha\rangle, |-\alpha\rangle, |-i\alpha\rangle\right\}$)~\cite{Becerra2013}. It was recently suggested that a PNRD receiver could also be useful for designing a nulling-based receiver to discriminate $M$-ary PSK signals at an error rate below the heterodyne detection limit~\cite{Izumi2013}. In addition, it should be noted that a similar technique is useful to reduce the discrimination error below what is possible with standard optical receivers, not only for phase modulated signals but also for intensity modulated signals such as on-off keying (OOK)~\cite{Cook2007,Tsujino2010} and pulse-position-modulation (PPM)~\cite{Guha2011,Chen2012}. Finally, a general design of a sequential-nulling receiver was recently proposed, which outperforms heterodyne detection for discriminating {\em any} $M$ spatio-temporal coherent-state waveforms by a factor of $4$ in the error-probability exponent~\cite{Nair2014}. These results on improved structured receivers for coherent state discrimination suggests that receivers based on photon counting could also be useful to go beyond the capacity limit of Gaussian receivers. This is because the task of a communication receiver is essentially to discriminate between $2^{nR}$ modulated codewords, each of which is a $n$-mode coherent-state waveform.

\begin{figure}
\begin{center}
\includegraphics[width=1\linewidth]{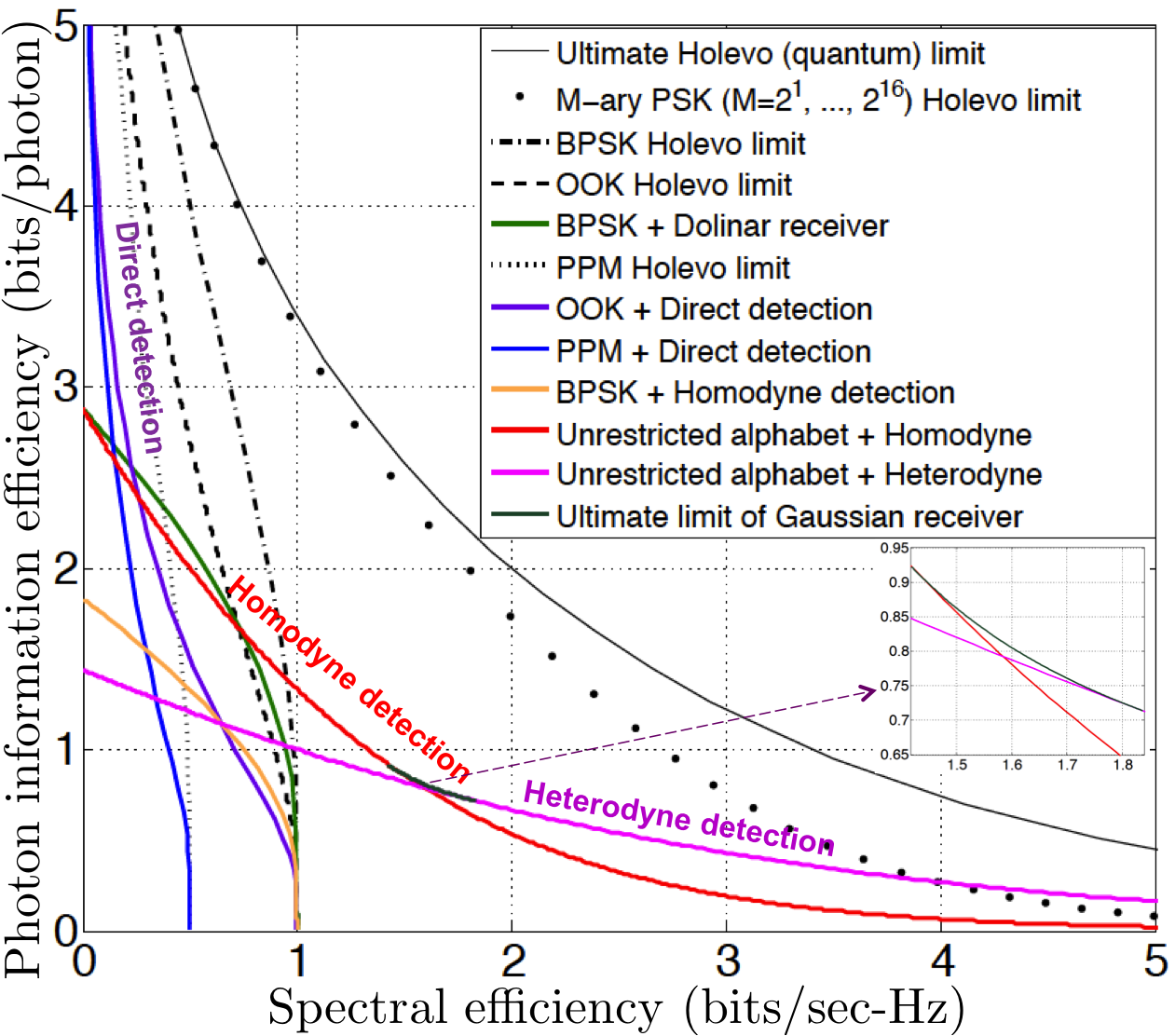}   %
\caption{(Color online) The tradeoff between photon information efficiency (PIE) and spectral efficiency (SE) for various choices of modulation formats and receivers. All non-black plots correspond to structured optical receivers, whereas the black lines are the Holevo capacities constrained to different modulation formats (i.e., with no restrictive assumption on the receiver). The thin black line on the top is the ultimate (Holevo) limit---no constraint on modulation and receiver---the highest capacity attainable over a pure-loss optical channel. }
\label{fig:non-Gaussian_receivers}
\end{center}
\end{figure}
For optical communication system designers, a popular way to assess the performance of a transceiver is to plot the trade-off between spectral efficiency (expressed in bits per symbol, or bits/sec/Hz) and the photon information efficiency (expressed in bits per received photon), for a given modulation format and a receiving strategy. For instance, for a pure loss channel with ${\bar n}$ mean received photon number per mode (or per time slot), using an optimal code and a Holevo-capacity-achieving receiver, the spectral efficiency (SE) is $g({\bar n})$ bits/sec/Hz and the photon information efficiency (PIE) is $g({\bar n})/{\bar n}$ bits/photon. When $\bar n$ is small, PIE is high and SE is small (this is the regime interesting for deep-space communication where every received photon is very precious), and in the high ${\bar n}$ regime, PIE is low and SE is high (this is the regime of interest for fiber-optic communication where the primary goal is to maximize the data rate). In Fig.~\ref{fig:non-Gaussian_receivers}, we plot the PIE-SE tradeoff for the Gaussian receiver limit (homodyne, heterodyne, and time-sharing between the two), the ultimate Holevo limit, and several different modulation and receiver strategies. For this plot, we chose the lossy optical channel ($N_{\rm th} = 0$) for simplicity.

In Fig.~\ref{fig:non-Gaussian_receivers}, all the lines plotted with non-black colors correspond to structured receivers, i.e., optical receivers the designs of are fully specifiable in terms of standard optical and electrical elements. The majority of plots in the figure pertain to discrete modulation formats (e.g., BPSK, OOK, and PPM). For computing the highest capacities attainable by coherent-detection (homodyne, heterodyne, and the optimal time-sharing between the two) receivers, as well as for evaluating the ultimate Holevo limit, we assume the optimal modulation, which for all those cases, is the continuous Gaussian modulation (i.e., when each symbol $|\alpha\rangle$ of a codeword is chosen i.i.d. from the distribution $p(\alpha) = \exp[-|\alpha|^2/{\bar n}]/\pi {\bar n}$, $\alpha \in {\mathbb C}$). 

The two non-Gaussian receivers we consider are single-photon detection (on-off direct detection), and the Dolinar receiver, a structured non-Gaussian receiver that can discriminate between any two coherent states at the minimum average error rate allowed by quantum mechanics~\cite{Dolinar1973,Cook2007}. The first observation to make is that in the low $\bar n$ (low SE, high PIE) regime, both the aforementioned non-Gaussian receivers---the Dolinar receiver (along with BPSK modulation) and the SPD receiver (along with either OOK or PPM modulation)---outperform the performance attainable by the general Gaussian receiver we studied in this paper. In the ${\bar n} \ll 1$ regime, the exact scaling of the Holevo limit ($C_{\rm ultimate}({\bar n}) = -{\bar n}\ln {\bar n} + {\bar n} + o({\bar n})$ nats/sec-Hz), and the capacity achievable by a single-photon-detection receiver ($C_{\rm SPD}({\bar n}) = -{\bar n}\ln {\bar n} - {\bar n}\ln \ln (1/{\bar n}) + O({\bar n})$ nats/sec-Hz) show that the gap between the two vanishes (i.e., their ratio goes to $1$) when ${\bar n} \to 0$~\cite{Chung2011}. Similarly, in the high $\bar n$ (low PIE, high SE) regime, the ratio of capacity attained by a coherent detection receiver (Heterodyne detection) and the Holevo limit goes to $1$ as ${\bar n} \to \infty$. Despite this, it is evident from Fig.~\ref{fig:non-Gaussian_receivers} that there is a substantial gap between the attainable performance by known structured receivers (all of which admit physical realizations via a symbol-by-symbol detection of the received modes) and the Holevo limit, even at moderate to high spectral efficiency. This gap in capacity is even more amplified when more than one spatial mode is employed, such as in a diffraction-limited near-field free-space optical channel~\cite{Guha2011a}. Even though some recent progress has been made on codes~\cite{Wil12} and receiver designs~\cite{Wil12a} that could in principle attain the Holevo limit, a fully structured, and a practically feasible, design of such a non-Gaussian optical receiver still eludes us. Finally, note that it is not just the receiver choice, but an appropriate choice of the modulation constellation commensurate with the photon number level, is important as well. Fig.~\ref{fig:non-Gaussian_receivers} shows that in the high photon number (high SE) regime, the capacity attained by heterodyne detection (with an optimally chosen modulation) becomes higher than the envelope of the Holevo rates of an $M$-ary PSK constellation (for $M = 2^1, 2^2, \ldots, 2^{16}$). This is not surprising since in the high photon number regime, the `circle' distribution does not approximate the circularly-symmetric Gaussian distribution $p(\alpha)$ well. This is why the envelope of the Holevo rates of $M$-ary PSK modulation is very close to the ultimate Holevo capacity at low $\bar n$ (since one circle in the phase space very well approximates the Gaussian) and peels off from it at higher $\bar n$ values.

\section{Conclusions}
\label{sec:conclusions}

The ultimate classical capacity of a lossy, noisy bosonic channel is attained by a transmitter that modulates coherent state (ideal laser-light) signals, albeit requiring a joint-detection receiver, which is hard to construct. In this paper, we restricted the receiver to a general {\em quantum Gaussian receiver}, which is made up of arbitrary quantum Gaussian operations (passive linear optics, squeezing and homodyne/heterodyne measurements) along with classical feedforward operations. We showed that the optimal Gaussian receiver strategy that maximizes the information rate is simply given by either homodyne or heterodyne detection, or time-sharing thereof. In other words, it was shown that any non-trivial Gaussian operation such as squeezing, partial measurements and conditional feedforward, do not help increase the communication performance over conventional homodyne and heterodyne detection receivers. In order to bridge the gap between the Shannon capacity 
limit of homodyne and/or heterodyne detection, and the ultimate Holevo capacity~\cite{Giovannetti2004}, the receiver must use non-Gaussian operations. We showed that in the low-photon flux regime, the direct detection receiver (a practical non-Gaussian receiver) as well as the Dolinar receiver (another structured non-Gaussian receiver) can outperform the capacity attained by Gaussian measurements. We quantified the gap between the Shannon capacity limits of all the above known structured optical receivers, and the Holevo limit---the maximum capacity attainable with any receiver structure permissible by physics---in terms of the trade-space between photon information efficiency (bits per received photon) and spectral efficiency (bits/sec-Hz). In order to attain the Holevo limit, the receiver must make collective measurements over long codeword blocks~\cite{Llo10,Guh11,Wil12,Wil12a}, which must include non-Gaussian elements (such as Kerr interactions, photon counting, or interactions with non-Gaussian states). Heralded realization of non-Gaussian states and measurements is an active area of theoretical and experimental study. An important theoretical question is to conceive of an experimentally feasible design of a non-Gaussian receiver that can attain the Holevo capacity, the maximum rate at which classical data can be reliably transmitted over an optical communication channel.

\begin{acknowledgements}
This work was performed while MT was on sabbatical from NICT at BBN. MT acknowledges kind hospitality of BBNers during his stay at BBN. SG was supported by the Defense Advanced Research Projects Agency's (DARPA) Information in a Photon (InPho) program, under Contract No. HR0011-10-C-0159.

\end{acknowledgements}

\appendix
\section{Characteristic functions and Gaussian states} 
\label{appendixA}

Here we briefly summarize the characteristic function formalism for Gaussian states and operations. More details can be found for instance in Refs.~\cite{Weedbrook2012, Giedke2002}. 

{\bf Characteristic function.}
Let us consider an $n$-mode bosonic system associated with 
an infinite-dimensional Hilbert space $\mathcal{H}^{\otimes n}$ 
and $N$ pairs of annihilation and creation operators, 
$\{\hat{a}_i,\, \hat{a}_i^\dagger\}_{i=1,\cdots,n}$, respectively, 
which satisfy the commutation relations
\begin{equation}
\label{eq:commutation_relation_a_a^dagger}
[ \hat{a}_i , \hat{a}_j^\dagger ] = \delta_{ij} .
\end{equation}
From these, one may construct the quadrature field operators:
\begin{eqnarray}
\label{eq:quadrature_x}
\hat{x}_i & = & \frac{1}{\sqrt{2}} ( \hat{a}_i^\dagger + \hat{a}_i ),\,{\text{and}}
\\
\label{eq:quadrature_p}
\hat{p}_i & = & \frac{i}{\sqrt{2}} ( \hat{a}_i^\dagger - \hat{a}_i ).
\end{eqnarray}
It is easy to verify that the commutation relations now translate to $[\hat{x}_i, \hat{p}_j]=i \delta_{ij}$. In the $n$-mode bosonic system, a quantum state with density operator $\rho$ is described by its characteristic function
\begin{equation}
\label{eq:characteristic_function}
\chi (x) = {\rm Tr} \left[ \rho \mathcal{W} (x) \right] ,
\end{equation}
where,
\begin{equation}
\label{eq:Weyl_op}
\mathcal{W} (x) = \exp \left[ - i x^T R \right] ,
\end{equation}
is a Weyl operator, 
$R=[ \hat{x}_1, \cdots , \hat{x}_n , \hat{p}_1, \cdots , \hat{p}_n]^T$ 
is a $2n$ vector consisting of quadrature operators, 
and $x =[ x_1, \cdots , x_{2n} ]$
is a $2n$ real vector. 
The overlap between any two operators $O_1$ and $O_2$ is
described by their characteristic functions as: 
\begin{equation}
\label{eq:overlap}
{\rm Tr}\left[ O_1 O_2 \right] = 
\left( \frac{1}{2\pi} \right)^n \int dx \, 
\chi_{O_1} (x) \chi_{O_2} (-x) .
\end{equation}

{\bf Gaussian states and operations}. 
The characteristic function for any Gaussian state is represented by 
\begin{equation}
\label{eq:chi_gaussian}
\chi (x) = \exp\left[ - \frac{1}{4} x^T \gamma x + i d^T x \right] ,
\end{equation}
where $2n \times 2n$ matrix $\gamma$ and $2n$ vector $d$ 
are called the covariance matrix and the displacement vector, respectively. 
Also, a Gaussian unitary operation is defined as a unitary operation 
that transforms Gaussian states to other Gaussian states. 
Any Gaussian unitary operation acting on a Gaussian state 
can be described by symplectic transformations 
of the covariance matrix and the displacement vector as 
\begin{equation}
\label{eq:symplectic}
\gamma \to S^T \gamma S, \quad d \to S^T d, 
\end{equation}
where $S$ is a symplectic matrix. For any covariance matrix, there exists a symplectic transformation that diagonalizes the covariance matrix (symplectic diagonalization). If the unitary operation includes only linear optical process (beamsplitters and phase shifts), then $S^T = S^{-1}$ and such a matrix $S$ is called an orthogonal symplectic matrix.

{\bf Decomposition of Gaussian unitary operation.} 
A symplectic matrix $S$ can always be decomposed as:
\begin{equation}
\label{eq:S_decomposition}
S = O \left( 
\begin{array}{cc}
M & 0 \\
0 & M^{-1}
\end{array}
\right) O' ,
\end{equation}
where $M$ is a positive diagonal matrix and $O$, $O'$ are orthogonal symplectic matrices. The physical meaning of the decomposition is that 
any Gaussian unitary circuit can be described by a sequential operation of linear optic circuit $O$, a product of single-mode squeezing operations, 
and another linear optic circuit $O'$. 

\section{Classical feedforward operation for a Gaussian state ensemble with a non-Gaussian distribution}
\label{appendixB}

In this appendix, we show that for a convex combination of Gaussian states, 
any trace decreasing Gaussian operation can be transformed into 
a trace preserving Gaussian operation. To this end, we use the characteristic functions whose basic properties were mentioned above in Appendix \ref{appendixA}.

\begin{figure}
\begin{center}
\includegraphics[width=1\linewidth]{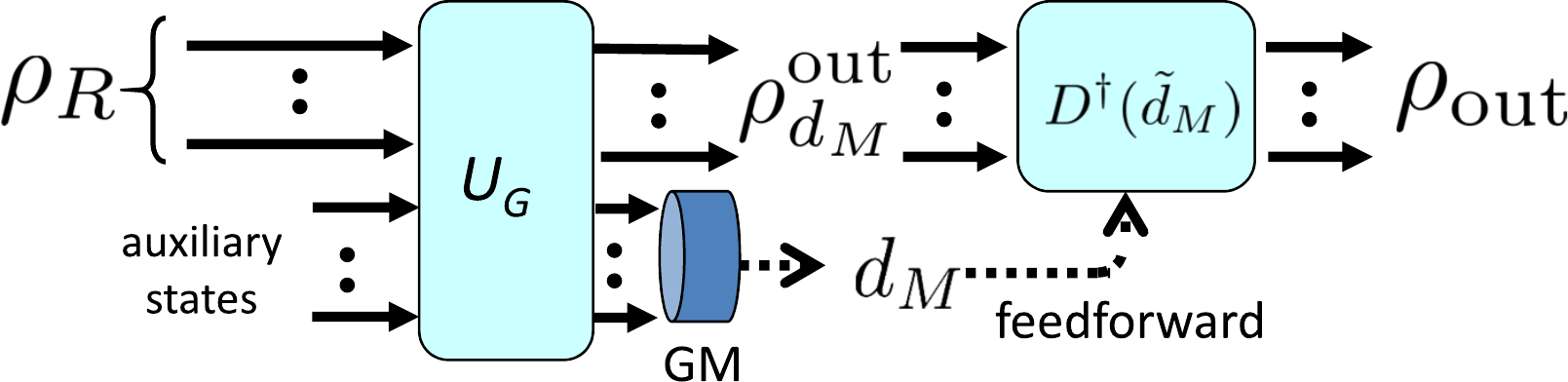}   %
\caption{\label{fig:single-step_ff}(Color online) 
Gaussian measurement with a single step feedforward. 
See the text for details.
}
\end{center}
\end{figure}

Recall the received state (Eq.~(\ref{eq:input_signal})) is given by  
$$
\rho_R = 
\int d^{2n} \underline{\alpha} P(\underline{\alpha}) 
\rho_r (\underline{\alpha}).
$$
Its characteristic function is given by a convex combination of 
the characteristic functions of $\rho_r(\underline{\alpha})$: 
\begin{eqnarray}
\label{eq:chi_R}
\chi_R(x) & = & {\rm Tr}\left[ \rho_R \mathcal{W}(x) \right]
\nonumber\\ & = & 
\int d^{2n} \underline{\alpha} P(\underline{\alpha}) 
{\rm Tr}\left[ \rho_r(\underline{\alpha}) \mathcal{W}(x) \right]
\nonumber\\ & = & 
\int d^{2n} \underline{\alpha} P(\underline{\alpha}) 
\chi_{\rho_r(\underline{\alpha})}(x) ,
\end{eqnarray}
where $\chi_{\rho_r(\underline{\alpha})}(x)$ is 
the characteristic function of $\rho_r(\underline{\alpha})$. 

Consider the trace decreasing operation consisting of a single step feedforward (FF) operation as illustrated in Fig.~\ref{fig:single-step_ff}, 
where an $n$-mode state $\rho_R$ is incident into 
an $n+m$-mode Gaussian unitary operation $U_G$ with 
an $m$-mode auxiliary Gaussian state $\rho_{\rm aux}$ 
and a part of the output (system $B$) is measured by an $m$-mode Gaussian 
measurement. Let ($\gamma_r$, $d_r$) and ($\gamma_{\rm aux}$, $d_{\rm aux}$) be sets of the covariance matrices and displacement vectors for  
$\rho_r(\underline{\alpha})$ and $\rho_{\rm aux}$, 
respectively. Let $S_G$ be a symplectic matrix for $U_G$. 
Then after operating $U_G$, the covariance matrix 
and displacement vector of the $n+m$-mode output are given by 
\begin{equation}
\label{eq:one-step_FF_U_G_output}
S_G^T (\gamma_r \oplus \gamma_{\rm aux}) S_G 
\equiv \left[ 
\begin{array}{cc}
A & C \\
C^T & B
\end{array}
\right] ,
\end{equation}
and,
\begin{equation}
\label{eq:one-step_FF_U_G_output_d}
S (d_r \oplus d_{\rm aux}) \equiv 
\left[
\begin{array}{c}
d_A \\ d_B
\end{array}
\right] .
\end{equation}
After measuring the system $B$ by a Gaussian measurement 
with the covariance matrix $\gamma_M$ and displacement 
(measurement outcome) $d_M$, we obtain the output in $A$ 
conditioned on $d_M$, whose covariance matrix and displacement 
are given by \cite{Giedke2002,Takeoka2008}:
\begin{equation}
\label{eq:U_G_out_cov}
\Gamma_{\rm out} = A - C^T \frac{1}{B+\gamma_M} C ,
\end{equation}
and, 
\begin{eqnarray}
\label{eq:U_G_out_displ}
d_{\rm out} & = & \left( d_A - C^T \frac{1}{B+\gamma_M} d_B \right) 
- C^T \frac{1}{B+\gamma_M} d_M 
\nonumber\\ & \equiv & 
\tilde{d}_{\rm out} + \tilde{d}_M 
\end{eqnarray}
where without loss of generality we assumed that $U_G$ 
does not include displacement operations (it can easily canceled by 
the inverse operation). Though this is a conditional operation on the measurement outcome $d_M$, one can easily eliminate $d_M$ in the output state by adding an additional displacement operation $D^\dagger(\tilde{d}_M)$ 
which results in the output state to be deterministically given by 
$\Gamma_{\rm out}$ and $d_{\rm out}$ those are independent of $\tilde{d}_M$. As a consequence, for input state $\rho_R$, we have the output: 
\begin{eqnarray}
\label{eq:chi_out}
\chi_{\rm out} = \int d^{2n} \underline{\alpha} 
P(\underline{\alpha}) \chi_{\underline{\alpha}} (x) ,
\end{eqnarray}
where $\chi_{\underline{\alpha}}(x)$ is a Gaussian characteristic function 
with $\Gamma_{\rm out}$ and $d_{\rm out}$. It does not include $d_M$ and thus independent of the partial measurement outcome. Thus the total operation is deterministic.


\begin{thebibliography}{99}
\bibitem{comment_P}
This does not lose the generality. Because 
if $P$ is nondiagonal, one can always diagonalize it at the beginning of 
the decoding process.
The diagonalization can be done by some linear optics circuit.
\end{thebibliography}
\end{document}